\documentstyle[11pt,paspconf,epsf]{article}

\begin{document}

\title{Clues to Outflows: Polarization of IRAS-Selected QSOs}
\author{Beverley J.~Wills}
\affil{Astronomy Department, RLM 15.308, University of Texas at Austin, Austin TX 78712}
\author{Dean C.~Hines}
\affil{Steward Observatory, University of Arizona, 933 N. Cherry Ave. Tucson, AZ 85721}

\begin{abstract}
Broad-band polarimetry and spectropolarimetry of a complete sample of 18 luminous AGNs
(L$_{bol} \ge 10^{11.5}$L$_{\sun}$), selected in an orientation-independent way by warm infrared flux,
shows that an orientation -- dust-covering Unified Scheme like that successfully explored for the
lower luminosity Seyfert galaxies applies to the luminous QSOs.  Broad emission lines and continuum seen
in polarized (scattered) light show that hyperluminous infrared galaxies contain buried QSO nuclei
and are therefore the `missing' QSO~2s -- the high-luminosity analogs of
Seyfert~2 galaxies like 
NGC~1068.  Most of the QSOs show significant polarization and reddening.  Three are `classical' 
low-ionization BAL~QSOs, and several others have narrower, blueshifted, low-ionization absorption.
These properties are to be compared with those of the UV-optically selected QSOs, of which only
$\sim$1\% show low-ionization BALs, and only the BAL~QSOs show significant polarization.  Outflows
are therefore much more common than previously suspected.
\end{abstract}


\section{The Relevance of Polarized IRAS AGNs to Mass Ejection from AGNs}
  The fraction of luminous AGNs showing blueshifted absorption  lines gives us the covering factor of
the outflowing gas --- but only if the sample is unbiased with respect to orientation.  Polarimetry and
spectropolarimetry allow us to detect buried QSOs in scattered light and thus to test 
orientation-dependent Unified Schemes for luminous AGNs, analogous to those for Seyfert nuclei.

The statistical relation of blue-shifted absorption lines to orientation\--depen\-dent obscuration and
scattering, and a comparison of line and continuum absorption in transmitted and scattered light paths in
individual objects, constrain the geometry and physical conditions in the absorbing gas.

%

We must first define an orientation-independent sample, and show that the QSO central
engines are actually the same kind, despite the different appearances at different orientations.

\subsection{How to Select an Orientation-Independent Sample}
\begin{figure}
\plottwo{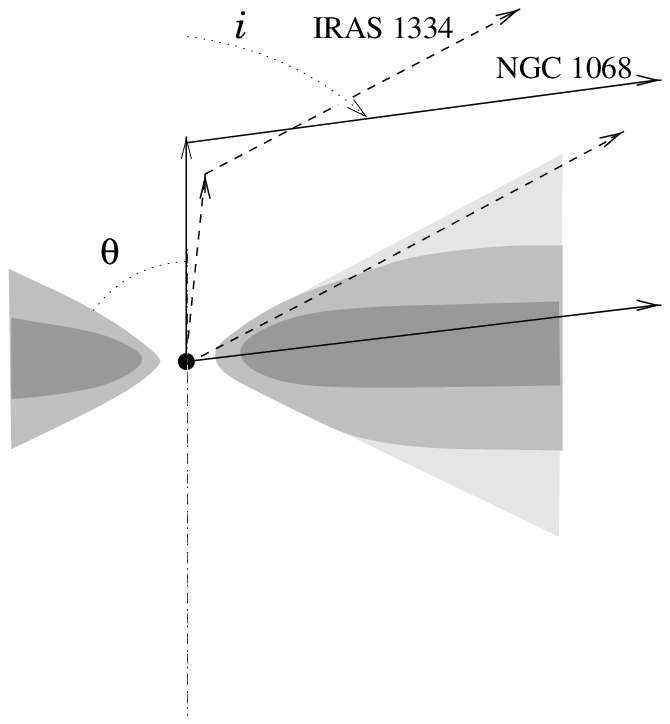}{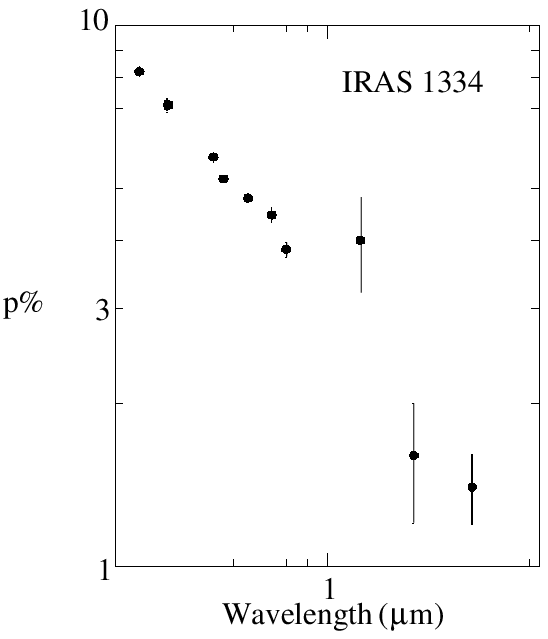}
\caption{(a, left) Cross-section through a dusty torus surrounding a continuum source and broad
line region (black dot).  $\theta$\ is the opening half-angle of the ionization or scattering
cone formed by the shadow of the torus, and {\it i} is the inclination 
of the line-of-sight to the 
axis (shown for NGC~1068).  Paths of scattered (polarized) and 
transmitted (reddened) light are shown 
for the buried Sy~1 in NGC~1068 (solid lines), and the less-obscured
QSO IRAS~13349$+$2438 (dashed lines).  \hfil\break
(b, right): Wavelength-dependent polarization of IRAS~13349$+$2438.}
\end{figure}

Unified Schemes successfully explain the diverse appearance of AGNs by means of different viewing
angles of an axisymmetric system.  Figure 1 illustrates the geometry deduced for the lower-luminosity
AGNs (Seyfert nuclei), and some luminous IRAS AGNs (Antonucci \& Miller 1985; Goodrich \& Miller 1994,
Wills et al. 1992; Hines \& Wills 1993).
The total light spectrum of NGC~1068 is dominated by starlight and a Seyfert~2 spectrum --
strong, unpolarized narrow lines viewed directly, arising within the ionization cone.  Viewed at
inclination {\it i}, equatorial extinctions of perhaps several 100 in A$_V$\ effectively block the
optical view of the central AGN ($\sim 10^m$\ at 20 \micron), but nuclear light scattered by
electrons within the opening angle of the torus reveals a Seyfert~1 continuum and broad emission lines in
polarized light.  In contrast, the first IRAS-discovered high-luminosity AGN (QSO) IRAS~13349$+$2438 is 
viewed at a smaller {\it i}.  The total light spectrum looks like that of a normal QSO, reddened, and
with very weak narrow-line emission.  However the percentage
polarization p\% is high and rises dramatically to the UV (Fig.~1b).  This is a result of a normal
unreddened QSO
spectrum seen in scattered, polarized light, diluted by
an unpolarized reddened spectrum of the QSO viewed directly, through absorbing dust.

UV-optical surveys are incomplete because they miss QSOs that lie within a dusty torus
or are buried in a dusty galaxy.  Despite high absorption along the line-of-sight, buried QSOs can be 
revealed in scattered (polarized) light.  Thus polarimetry and spectropolarimetry allow us to test 
Unified Schemes for luminous AGNs (QSOs) -- to show whether the dusty hyperluminous infrared galaxies
(HIGs) contain the same kind of active nucleus as the UV-optically selected QSOs.

The optically thick torus may intercept an appreciable fraction of the QSO's UV-optical
luminosity.  At a few pc from the QSO nucleus, dust grains will be heated to an equilibrium
temperature of 1000 K, radiating near 3\micron, and at $\sim$1 kpc, grains at 100 K will radiate
near 30 $\mu$m.  The dusty gas may generally be optically thin at mid-IR wavelengths.  This allows
an approximately orientation-independent sample to be selected by warm IR colors.

\section{The Sample and Observations}
  The sample is shown in Table~1.  The 12 QSOs were derived from the complete sample defined
by Low et al. (1988) who selected sources from the IRAS Point Source Catalog with 
$|$b$| > 30\deg$, detected at 60 $\mu$m, and with F$_{\nu}$(25$\mu$m)/F$_{\nu}$(60$\mu$m) $>$1/4.
They present only the broad-emission line objects.  We have selected those with
L(IR) $> 10^{11.5}$L$_{\sun}$\ (H$_0$ = 100 km s$^{-1}$ Mpc$^{-1}$).  We excluded radio
core-dominant objects because their spectra are contaminated by
synchrotron emission.  We included two additional QSOs at the flux limit of Low et al.'s
sample.  From the literature, we added all 6 objects known to satisfy the same
luminosity, flux, and IR color criteria as the Low et al. QSOs.  
All 6 show strong narrow-line emission.  This is likely to be a lower
limit to the true number of these HIGs.

We measured linear polarization in unfiltered light for the whole sample, using the Breger
broad-band polarimeter on the McDonald 2.1-m Struve telescope. 
Objects showing significant p\% were measured through filters -- CuSO$_4$,
RG630, and standard UBVRI bands.
Spectropolarimetry for 3 QSOs and 3 HIGs was obtained using
the LCS spectropolarimeter on the McDonald 2.7-m Smith telescope.  Table 1 summarizes our
results.  Further details of our
measurements, and data from other sources, may be found in Wills \& Hines (in preparation),
Hines (1994), and the references to Table~1.

\begin{table}
\caption{Summary of Polarimetry Results\tablenotemark{a}} \label{tbl-1}
\begin{center}\scriptsize
\begin{tabular}{lllrllll}
Name & log L$_{bol}\tablenotemark{b}$ & Class & p$_{max}$\%\tablenotemark{c} & p$_{\lambda}$\tablenotemark{d} & Absorbing & Polarized & New\\
     &                                &       &                              &                                & Outflow\tablenotemark{e} & Compnt\tablenotemark{f} & Class\\
\tableline
\\
IRAS 05189$-$2524  &11.9  &HIG &5.0$\pm$0.1  &\bf +  &$\cdots$   & BLR,cont & QSO 2\\
IRAS 20460$+$1925  &12.9  &HIG &5.0\ \ 0.5  &\bf +  &$\cdots$    & BLR,cont & QSO 2\\
IRAS 23060$+$0505  &12.5  &HIG &8.1\ \ 0.3  &\bf +  &$\cdots$    & BLR,cont & QSO 2\\
\\
IRAS 09104$+$4109  &12.6  &HIG &21.0\ \ 3.2 &\bf $-$  &$\cdots$    & BLR,cont & QSO 2\\
IRAS F10214$+$4724 &14.4  &HIG &18\ \ \ \ \ldots &\bf $-$ &$\cdots$    & BLR,cont & QSO 2\\
Mkn 463 E        &11.8  &HIG &13\ \ \ \ \ldots &\bf $-$  &$\cdots$    & BLR,cont & QSO 2\\
\\
I Zw 1           &12.0  &QSO   &1.7\ \ 0.2  &\bf $-$  &LOAL?      & ?,cont & QSO 1\\
Mkn 231          &12.4  &QSO   &10.5\ \ 0.3 &\bf $-$  &LOBAL,LOAL & BLR,cont & QSO 1\\
IRAS 13349$+$2438  &11.6  &QSO   &8.0\ \ 0.2  &\bf $-$  &LOAL?      & BLR,cont & QSO 1\\
IRAS 14026$+$4341  &12.1  &QSO   &12.0\ \ 0.7 &\bf $-$  &LOBAL,LOAL & BLR,cont & QSO 1\\
\hfil(CSO 409) &\\
\\
IRAS 00275$-$2859  &12.7  &QSO   &1.5\ \ 0.1  &\bf 0  &$\cdots$ & $\cdots$ & QSO 1\\
IRAS 07598$+$6508  &11.8  &QSO   &2.2\ \ 0.1  &\bf 0  &LOBAL,LOAL & cont  & QSO 1\\
PG 1700$+$518      &11.9  &QSO   &0.5\ \ 0.1  &\bf $-$ &LOBAL,LOAL & ?,cont  & QSO 1\\
IRAS 21219$-$1757  &11.8  &QSO   &1.5\ \ 0.2  &\bf $-$ &$\cdots$ & ?,cont  & QSO 1\\ 
\\
PHL 1092         &11.9  &QSO   &0.3\ \ 0.3  &$\cdots$    &$\cdots$    & $\cdots$ & QSO 1\\ 
Mkn 1014         &12.5  &QSO   &0.7\ \ 0.2  &$\cdots$    &$\cdots$    & $\cdots$ & QSO 1\\ 
IRAS 04505$-$2958  &12.0  &QSO   &0.4\ \ 0.2  &$\cdots$    &$\cdots$    & $\cdots$ & QSO 1\\ 
IRAS 13218$+$0552  &11.7  &QSO   &0.7\ \ 0.4  &$\cdots$    &$\cdots$    & $\cdots$ & QSO 1\\ 

\end{tabular}
\end{center}


\tablenotetext{a}{The groups of objects are arranged in order of decreasing
obscuration, or, on a model as in Fig.~1, decreasing inclination of the line-of-sight to the
axis of symmetry, assuming that such an axis of symmetry can be defined for these
QSOs, as it is in many Seyfert galaxies.}
\tablenotetext{b}{The bolometric luminosity is in units of L$_{\sun}$, assuming
H$_0 = 100$ km s$^{-1}$ Mpc$^{-1}$.}
\tablenotetext{c}{This gives the maximum observed polarization from
broad-band data or spectropolarimetry.}
\tablenotetext{d}{The wavelength dependence of polarization.  The `{\bf +}',
`{\bf $-$}', and `{\bf 0}' symbols refer to polarization increasing with
wavelength, decreasing, or with little wavelength dependence, respectively.
In some cases the observed p\% is too small for the wavelength dependence to be defined.}
\tablenotetext{e}{`Absorbing Outflow' refers to the existence of blueshifted
low-ionization absorption lines -- either broad (LOBAL), or narrow (LOAL,
Boroson \& Meyers 1992 \apj, 397, 442).}
\tablenotetext{f}{`Polarized Component' indicates which is seen in
polarized flux -- either the BLR or continuum.}
\tablenotetext{}{References to polarization data are:\\
Berriman, G. , Schmidt, G. D.  \& West, S. C. 1990, \apjs, 74, 869 (Mkn 1014)\\
Hines, D., Schmidt, G., Smith, P., \& Weymann, R. 1995, \baas, 187, 8409 (PG1700+518)\\
Hines, D. C. \& Wills, B. J. 1993, \apj, 415, 82 (IRAS 09104+4109)\\
Hines, D. C. \& Wills, B. J. 1995, \apjl, 448, 69L (IRAS 07598+6508)\\
Jannuzi, B. T., Elston, R., Schmidt, G. D., Smith, P. S., \& Stockman, H. S.  1994, \apjl, 429,\\
49 (IRAS F10214+4724)\\
Kay, L. E.  1994, \apj, 430, 196 (Mkn 463E)\\
Miller, J. S. \& Goodrich, R. W.  1990, \apj, 355, 456 (Mkn 463E)\\
Tremonti, C. A., Uomoto, A., Antonucci, R. R. J., Tsvetanov, Z. I., Ford, H. C., \& Kriss, G.\\ 
A.  1996, \baas, 189, 1105 (Mkn 463E)\\
Smith, P. S., Schmidt, G. D., Allen, R. G., \& Angel, J. R. P.  1995, \apj, 444, 146 (Mkn 231)\\
Stockman H.S., Moore R.L., \& Angel J.R.P.
1984, \apj, 279, 485 (PHL 1092)\\
Webb, W., Malkan, M., Schmidt, G., \& Impey, C.  1993, \apj, 419, 494 (I Zw 1)\\
Young, S., Hough, J. H., Efstathiou, A., Wills, B. J., Bailey, J. A., Ward, M. J., \& Axon, D.\\ 
J.  1996, \mnras, 281, 1206 (Mkn 463E, Mkn 231)\\
Wills, B. J., Wills, D., Evans, N. J., Natta, A., Thompson, K. L., Breger, M., \& Sitko, M. L.\\
1992, \apj, 400, 96 (IRAS 13349+2438)
}

\end{table}

\section{Results}
The sample AGNs are grouped in Table~1 in order of decreasing obscuration.  On the axisymmetric dusty
torus model
of Fig.~1 this could be interpreted as decreasing inclination.   The spectra of the first two groups
(HIGs) are
dominated by starlight and strong, Seyfert~2-like narrow lines.  When corrected for starlight
p\% is high (up to 26\%).  The polarized flux spectra show reddened QSO
continuum and BLR.  In the first group the scattered light is even redder than starlight, resulting in a 
characteristic increase in p\% to longer wavelengths.  In contrast, in the second group, the scattered
QSO continuum and broad lines are much less reddened, resulting in higher observed p\%
increasing towards the UV.  All the remaining groups show QSO broad lines and continuum in total and
direct light.
In the third group the total spectrum is reddened but p\% increases to the UV -- a result of a reddened direct
view of the nucleus but much less reddened scattered spectrum.  p\% is generally quite high.  Narrow
associated absorption or weak, broad absorption is seen in all these objects.  The objects of the fourth
group all have significant but low p\% and at least 3 show low-ionization narrow and broad associated 
absorption.  These spectra are less reddened than the preceding groups.  At least in IRAS~07598$+$6508
the broad emission lines are unpolarized.  Taking into account spectropolarimetry of other
BAL~QSOs as well, apparently a lower inclination view of the nucleus results in multiple (direct and
scattered)
views of the QSO nucleus.  Finally, the QSOs of the fifth group appear as unpolarized, like nearly all of 
the 114 UV-optically selected PG~QSOs (Berriman et al. 1990).  On an axisymmetric torus model, these are
relatively unobscured QSOs, viewed within the opening angle ({\it i} $< \theta$).

\section{Interpretation}

Broad-band polarimetry and spectropolarimetry of 12 QSOs and 6 HIGs selected by warm 60\micron\ and
25\micron\ flux densities, and therefore essentially unbiased with respect to orientation, shows that
these are analogous to the Seyfert~1 and 2 classes at lower luminosities.  Spectropolarimetry reveals
QSO-like polarized broad emission lines and continua in all 6 HIGs.  This means that a dust-orientation
Unified
Scheme holds for luminous AGNs, and the HIGs -- the `missing' QSO~2s -- 
constitute $\sim$30\% of this population.

\begin{figure}
\plotfiddle{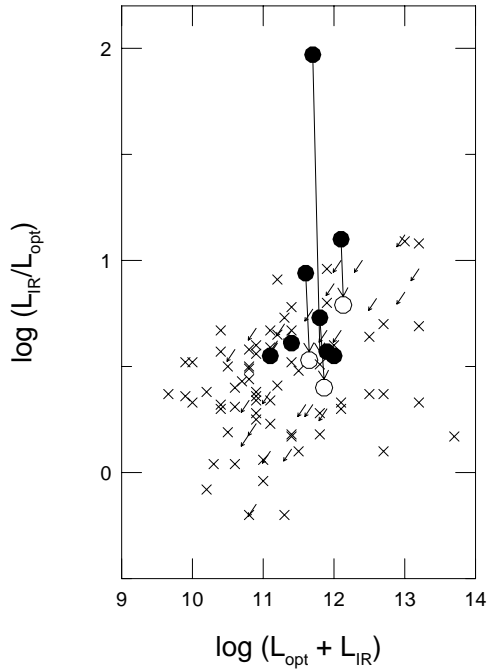}{8.4cm}{0}{50}{50}{-130}{-20}
\caption{A comparison of log L$_{\rm IR}$/L$_{\rm opt}$ for IRAS-selected QSOs ($\bullet$), with optically
selected PG QSOs ($\times$, and upper limits) of the same bolometric luminosity.
Reddening corrected values ($\circ$), show that IR-selected QSOs do not
have enhanced L$_{\rm IR}$\ compared with optically-selected QSOs.  Figure adapted from Low et al. (1989).
} 
\end{figure}

The axisymmetric torus model is suggested by analogy with Seyfert nuclei where radio-jets, ionization
cones or dust disks define the axis, and, for IRAS 13349$+$2438, by the major axis of its host galaxy.
However, the sequence we have described could be simply one of decreasing dust content.  If dust 
content were greater for the more reddened objects, we would expect their L$_{\rm IR}$\ to be enhanced.
A preliminary test of this is shown in Fig.~2 where we compare the ratio L$_{\rm IR}$/L$_{\rm opt}$\
for the Low et al. QSOs with that for the optically selected PG QSOs of the same bolometric luminosity.
When corrected for line-of-sight reddening, L$_{\rm IR}$/L$_{\rm opt}$\ ratios are the same for the IR-
and optically selected QSOs.

While the numbers are small and the number of HIGs is uncertain, from the different views at different
inclinations we build a picture of a typical QSO as follows: $\sim$30\% (6 HIGs) of the center is covered
by thick dust, $\sim$40 -- 45\% by thinner dust (groups 3 \& 4), with 20 -- 25\% of the QSO relatively
unobscured (the 5th group).  Blueshifted absorption lines are associated with partially obscured views
of the nucleus.  Only 3-4 QSOs are sufficiently unobscured to have been selected in UV-optical QSO
surveys, so the true space density of QSOs has been underestimated by a factor $\sim $5.  
Webster et al. (1995) estimate that a similar fraction of radio-loud 
quasars (80\%) have been missed as a result of reddening.

On the axisymmetric torus model we can assign a relatively clear view within a cone of 
$\theta \sim 22\deg$, a partially obscured zone with low-ionization absorbing outflows between 
$\theta \sim 22\deg$\ and $50\deg$,
and a thick dusty torus with $\theta \sim 50\deg$.  These numbers are consistent
with half-opening angles deduced from spectropolarimetry of individual objects of the sample (see
references in Table~1).

Low-ionization absorbing outflows are related to dusty gas -- neutral and warm (see Grupe et al.,
Mathur, this volume) -- and, whichever dust\--orienta\-tion Unified Scheme holds,
they are much more common than previously thought\footnote
{In related work, Hines \& Schmidt (these Proceedings) use polarization of a large sample of BAL~QSOs
to investigate their relation to HIGs.}.


\acknowledgments
We thank D. Wills, M. Breger, J. H. Hough, R. W. Goodrich, D. R. Doss, E. Dutchover and V. Vats 
(Karl Lambrecht Corporation) for observing and instrumentation help, and NASA (grant NAG5-3431).

\begin{question}{Kirk Korista}
Are all the BAL QSOs in your sample Mg\,II-type BAL QSOs?  Would you know whether your sample included
high-ionization (``normal'') BAL QSOs?
\end{question}
\begin{answer}{Bev Wills}
The sample QSOs, being IRAS-selected, are all low redshift, so any information on the (UV)
high-ionization BALs must come from IUE or HST spectroscopy.  In all cases for which that information
is available,
objects with low-ionization BALs also show high-ionization BALs.  I do not know any
exceptions to this.  Apart from the objects in our sample that we know have low-ionization 
absorption, I don't think that UV spectroscopy is available.  UV spectroscopy would provide
an important clue to the
relation between low-ionization and high-ionization-only broad absorption line gas, with or
without axisymmetric torus models (see Weymann, Morris, Foltz, \& Hewett 1991, \apj, 373, 23).
\end{answer}

\end{document}